\documentclass[pre,aps]{revtex4}
\usepackage{graphicx}
\newcommand{\til}{\tilde t}

\begin{document}
\title{Physical ageing studied by a device allowing for rapid thermal equilibration}
\author{Niels Boye Olsen, Tina Hecksher, Kristine Niss and Jeppe C. Dyre}
\affiliation{DNRF Centre ``Glass and time'', IMFUFA,
Department of Sciences, Roskilde University, Postbox 260, DK-4000 Roskilde, Denmark}
\date{\today}

\begin{abstract}
Ageing of organic glasses to the equilibrium liquid state is studied by measuring the dielectric loss utilizing a microregulator where temperature is controlled by means of a Peltier element. Compared to conventional equipment the new device adds almost two orders of magnitude to the span of observable ageing times. Data for five organic glass-forming liquids are presented. The existence of an ``inner clock'' is confirmed by a model-free test showing that the ageing of structure is controlled by the same material time that controls the dielectric properties. At long times relaxation is  not stretched, but simple exponential, and there is no ``expansion gap'' between the limits of the relaxation rates following up and down jumps to the same temperature.
\end{abstract}


\maketitle

\section{Introduction}

Ageing is the change of materials properties over time. Ageing phenomena often involve chemical degradation, but there are also many cases of purely physical property changes. A well-known example is the slight shrinking of plastic rulers over years deriving from an extremely slow density increase with time. Understanding physical ageing is obviously important for applications, but physical ageing also presents fundamental scientific challenges and provides valuable insight into materials properties. This paper shows that utilizing the Peltier thermoelectric effect ageing studies may be extended to considerably shorter times than have so far been possible, adding almost two decades to the span of observable ageing times.

A prime example of physical ageing is that of a viscous liquid's properties relaxing close to its glass transition. In equilibrium a liquid's properties do not change with time, of course. If temperature is changed, properties gradually adjust themselves to new equilibrium values. If temperature is lowered, a glass is produced; recall that by definition a glass is nothing but a highly viscous liquid that has not yet had time to relax fully to equilibrium \cite{kau48,deb96,ang00,alb01,kob04,dyr06}. A glass gradually approaches the equilibrium liquid state. This state that can only be reached on laboratory time scales, however, if the system is kept just below its glass transition temperature (``annealed'') -- contrary to the popular myth windows do not flow observably.

Ageing is an intrinsically nonlinear phenomenon. This is because the ageing rate is structure dependent and itself ages with time 
\cite{kov63,moy76,maz77,str78,mck94,hod95,avr96}. Therefore ageing studies provide information beyond those obtained by linear-response experiments like dielectric relaxation measurements. A simple ageing experiment consists of a temperature step, i.e., a rapid decrease or increase of temperature to a new, constant value. Ideally the new temperature should be well-defined homogeneously throughout the sample and constant in time before any relaxation has taken place. If that is achieved, it is possible to monitor the complete relaxation to equilibrium of the physical property that is being probed.

\section{Experimental}

What are the requirements for such an ideal ageing experiment? First, one should obviously have very good temperature control. Secondly, a physical observable is needed that may be monitored fast and accurately and which -- preferably -- changes significantly for small temperature changes. Third, the set-up should allow for fast temperature changes. 

As regards the last point, in current state-of-the-art ageing experiments the characteristic thermal equilibration time $\tau$ is at least 100 s ($\tau$ is defined by writing the long-time approach to equilibrium $\propto\exp(-t/\tau)$). This reflects the fact that heat conduction is a notoriously slow process. Experience shows that in order to monitor a virtually complete ageing curve for a temperature down jump, at least four decades of time must be covered. Thus with present methods, in the best cases one needs of order $100\,{\rm s}\times 10^4 = 10^6\,{\rm s}$ to perform an ideal temperature down-jump experiment. This is more than a week. 

\begin{figure}
  \includegraphics[width=12cm]{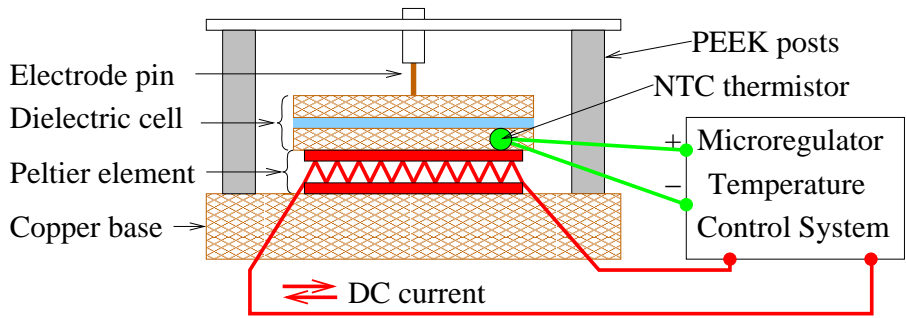}
  \caption{Schematic drawing of the dielectric measuring cell with the microregulator. Liquid samples are deposited in a 50 $\mu$m gap between disks of the dielectric cell. The Peltier element either heats or cools the dielectric cell, depending on the direction of the electrical current powering the element. The current is controlled by an analog temperature control system that receives temperature feedback information from an NTC thermistor embedded in one disk of the dielectric cell. A stainless steel electrode pin  keeps the cell pressed against the Peltier element and provides the electrical connection to one of the disks.} 
\end{figure}

Feynman is famous for his prophesy of the nano-revolution by remarking that there is ``plenty of room at the bottom.'' For ageing experiments there is also plenty of  ``room at the bottom'', but here at the bottom of the time axis: Clearly much is to be gained if it were possible to equilibrate sample temperatures much faster. In order to be able to make faster temperature-jump experiments we have designed a dielectric cell based on a Peltier thermoelectric element by means of which heat flow is controlled via electrical currents (Fig. 1). The characteristic thermal equilibration time of the new microregulator is two seconds \cite{iga08}. This is 50 times faster than the best conventional equipment. If required, temperature may be kept constant over weeks, with fluctuations smaller than 100 $\mu$K.

For monitoring ageing, measuring the dielectric loss at some fixed frequency is probably the best choice. With modern equipment this quantity may be monitored fast and accurately. Moreover, for viscous liquids in the right frequency range the dielectric loss changes considerably for small temperature variations. The use of dielectric relaxation measurements for monitoring ageing was pioneered by Leheny {\it et al.} \cite{leh98,yar03} and by Lunkenheimer {\it et al.} \cite{lun05}.

Using the microregulator we studied ageing of dibutyl phthalate (DBP), a widely studied glass-forming organic liquid with glass transition around 176 K. Ageing was studied by monitoring how the dielectric loss, $\epsilon''(f=0.37{\rm Hz})$, develops as a function of time after four 2K temperature steps starting from equilibrium at $T=172, 174, 176$ K. In order to ensure equilibrium before each temperature jump, the sample was kept at that particular temperature for so long that there were eventually no detectable changes of the dielectric properties. Figure 2(a) shows the results, jumping from $172$ K and $176$ K, respectively, to $174$ K -- as well as the opposite jumps. The inset shows a blow-up of the final stages of approach to equilibrium at $174$ K. Note that ageing for jumps to a given temperature is considerably faster when starting at a higher temperature than when starting at a lower (the two jumps to $174$ K, in the middle of Fig. 2(a)). This is the so-called fictive-temperature effect described already by Tool in the 1940's \cite{too46}. 

\begin{figure}
  \includegraphics[width=6.5cm]{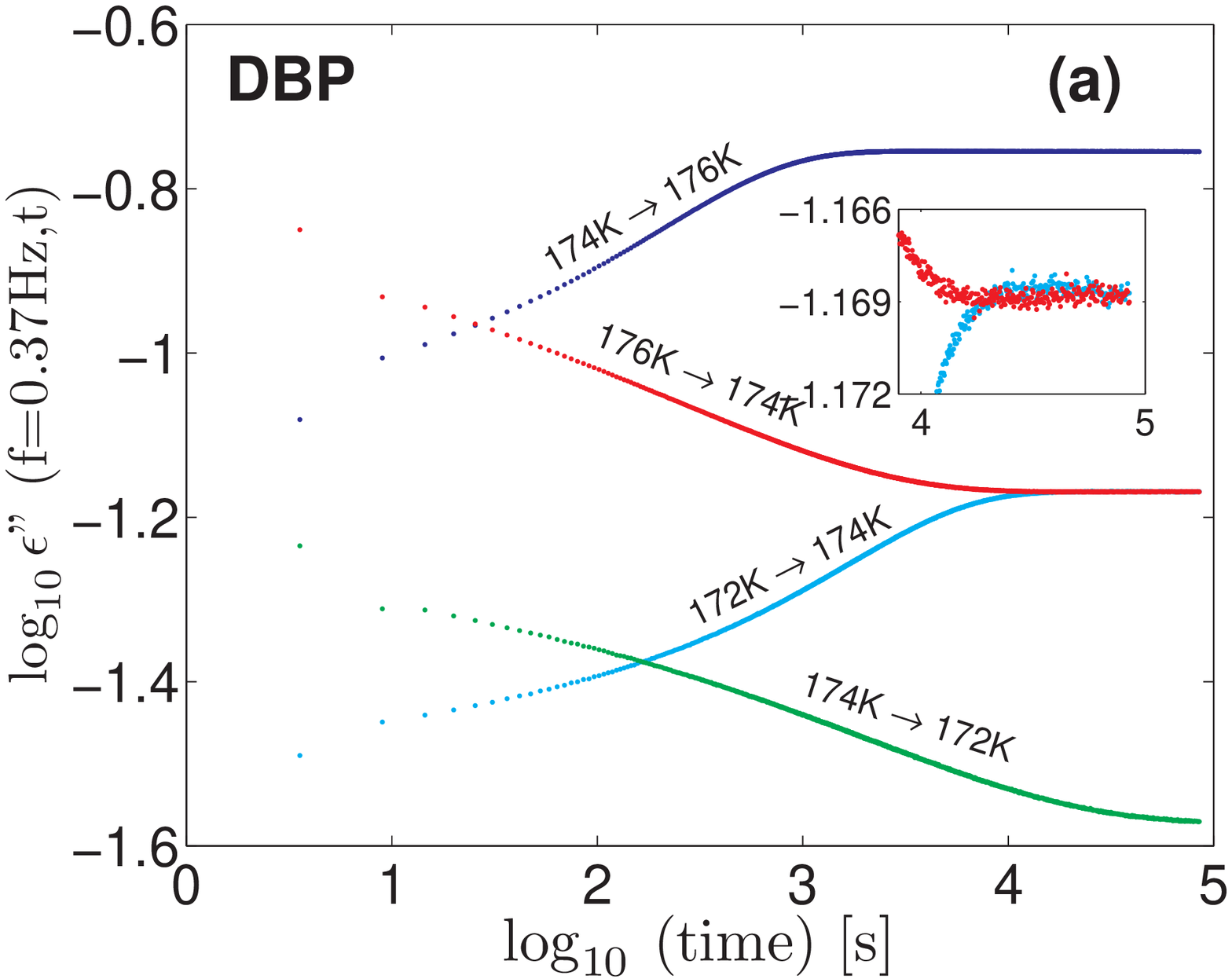}
  \includegraphics[width=6.5cm]{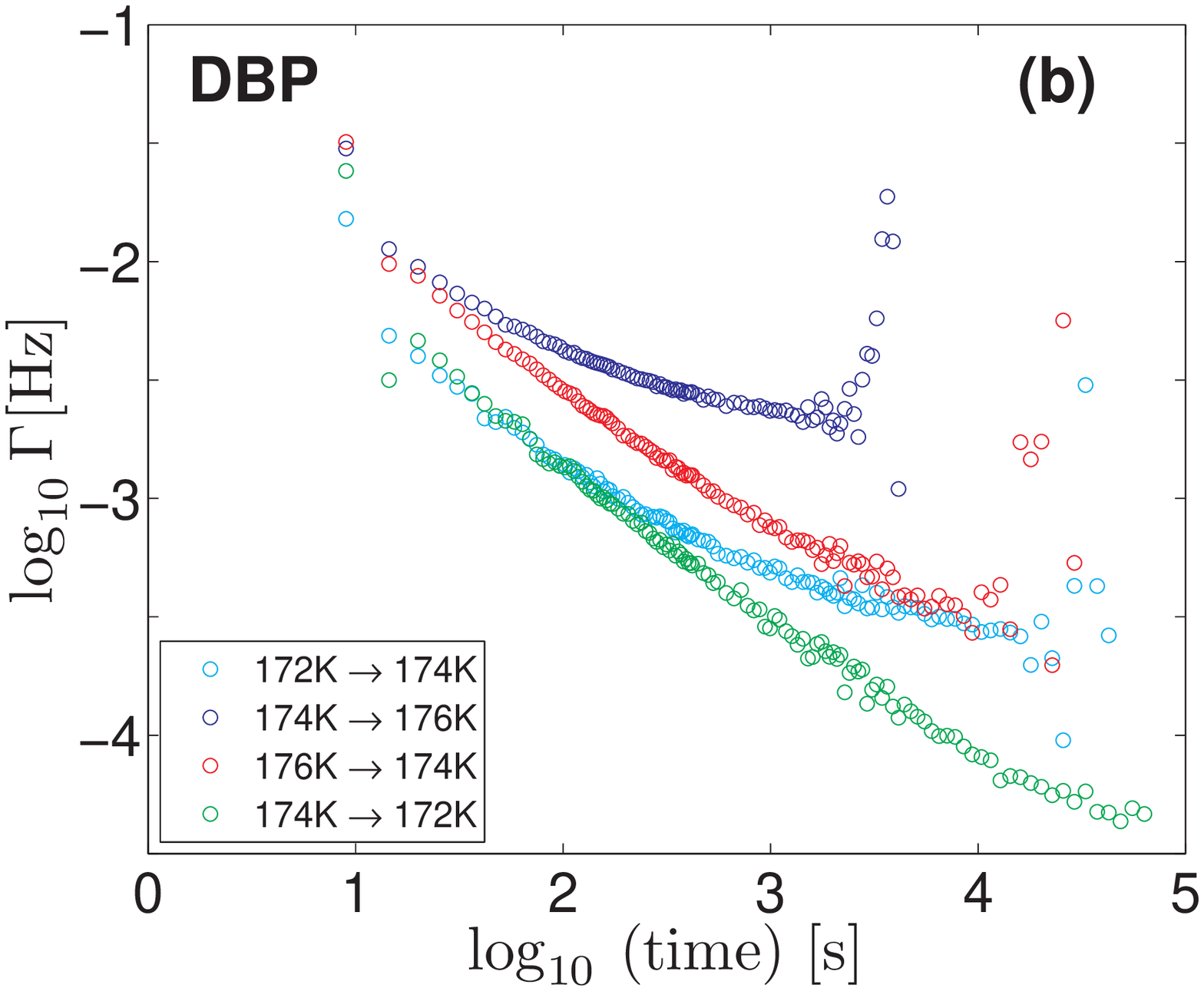} 
  \includegraphics[width=6.5cm]{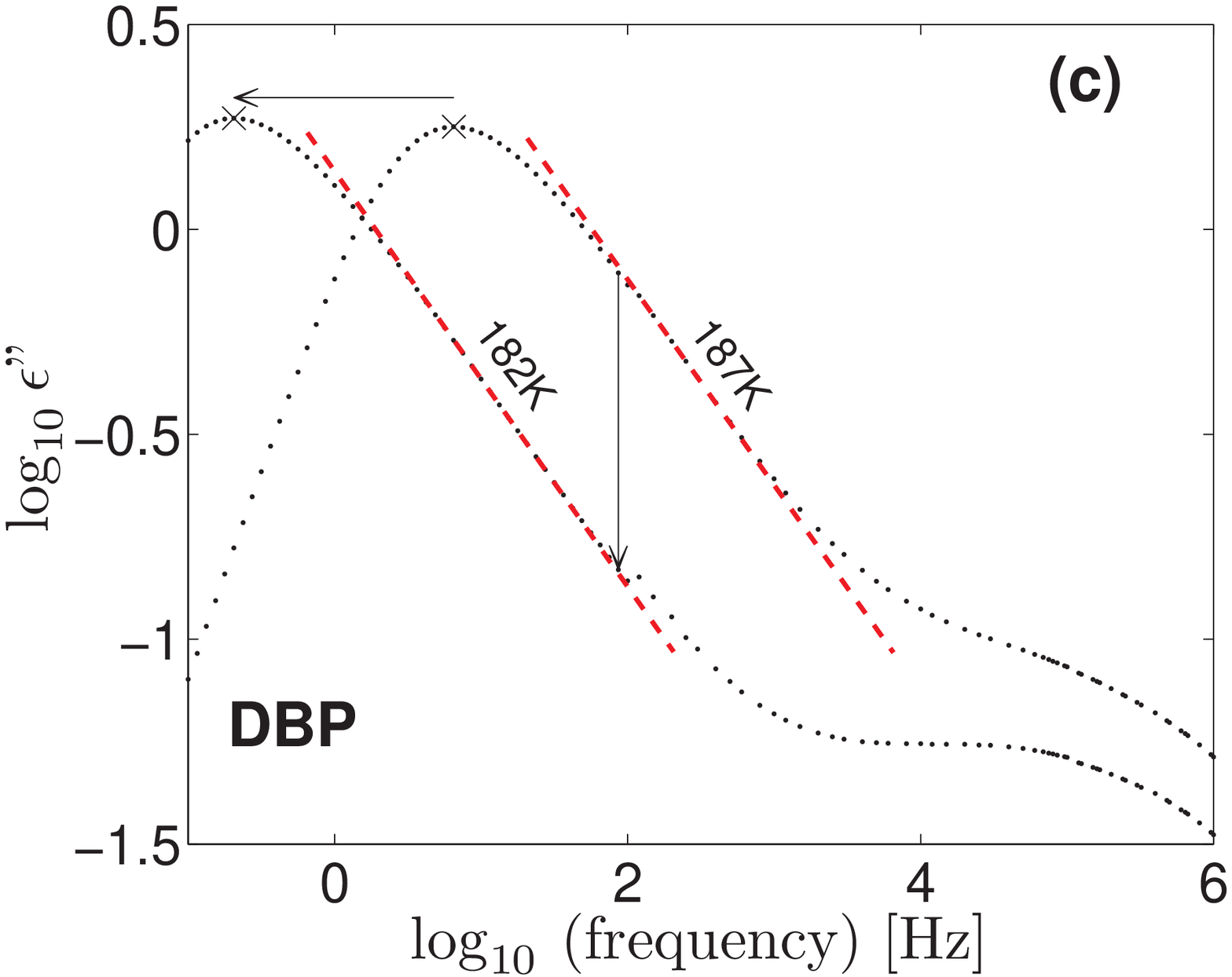} 
  \includegraphics[width=6.5cm]{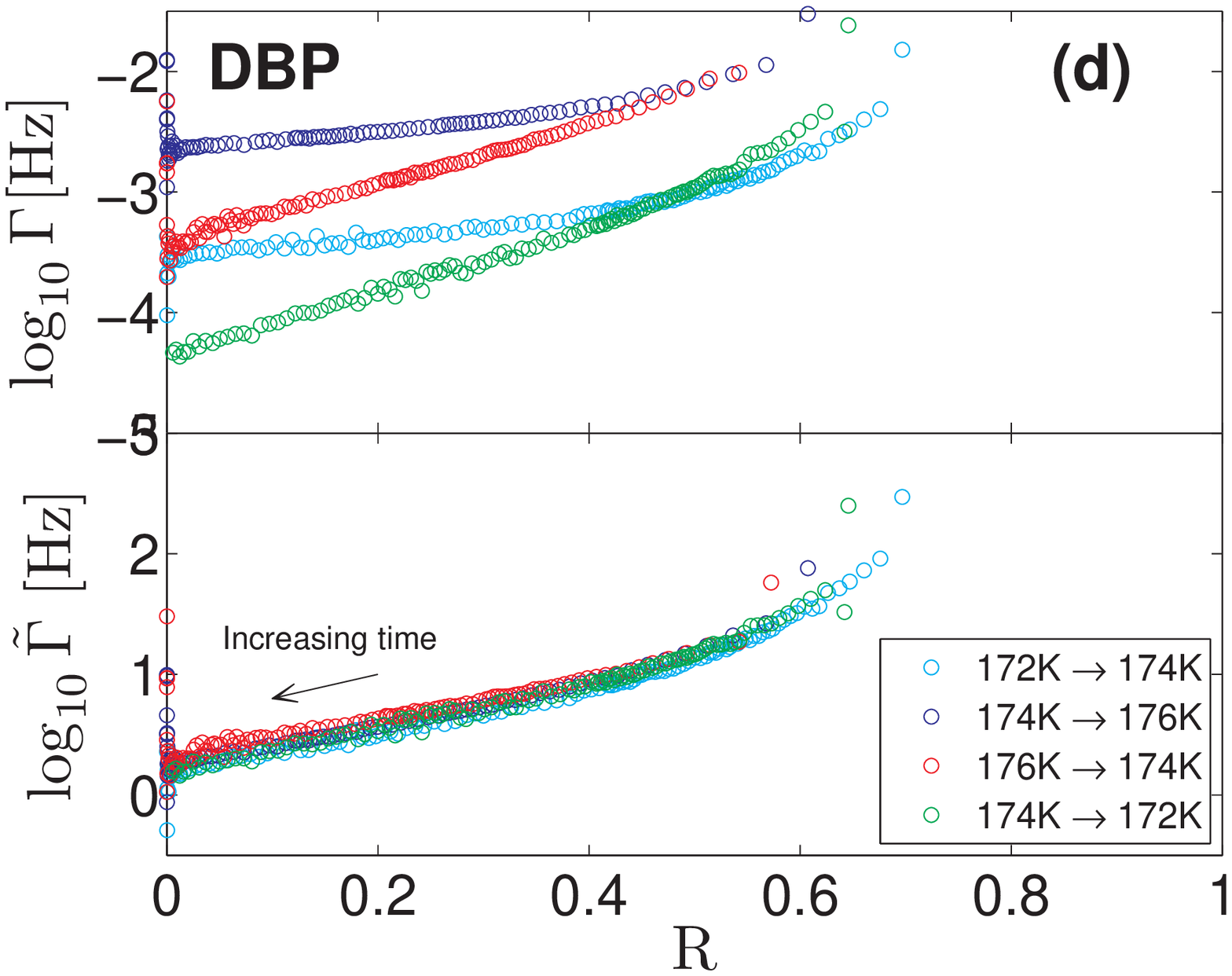}
 \caption{Ageing of dibutyl phthalate (DBP) studied by monitoring the dielectric loss at 0.37 Hz. (a) Ageing following 2K temperature jumps from $172,174,176$ K. The inset shows the final stages of approach to equilibrium at $174$ K from the two jumps from $172$K and $176$ K. (b) The Kovacs-McKenna relaxation rate $\Gamma$ defined by eq. (\ref{KM}) for the same data. Up and down jumps to $174$ K at long times give same relaxation rate, which shows that there is no ``expansion gap'' as proposed by Kovacs in 1963 \cite{kov63}. (c) Dielectric loss at equilibrium conditions at two higher temperatures where the loss peak is within the measured frequency window. This figure illustrates how one utilizes the fact that the loss varies as $f^{-1/2}$ at high frequencies to measure the dielectric loss peak frequency giving the inner clock's clock rate. Details are provided in the supplementary information. (d) The KM relaxation rate $\Gamma$ (top panel) and its dimensionless version $\tilde\Gamma$ (bottom panel,
defined in the text eq. \ref{mat_time_def}-\ref{gammadim_def}) plotted as functions of the normalized relaxation function. The data collapse in the second figure confirms the existence of an inner clock (a material time).}
\end{figure}

Quite generally a relaxation function $r(t)$ is defined by subtracting the long-time limit of the physical property monitored. In our case:  

\begin{equation}\label{relax_fct}
r(t)\,=\,|\ln\epsilon''(f=0.37{\rm Hz},t)-\ln\epsilon''(f=0.37{\rm
  Hz},t\rightarrow\infty)|\,. 
\end{equation}
For any relaxation function the Kovacs-McKenna (KM) relaxation rate
$\Gamma$ is defined \cite{kov63,mck94} by
\begin{equation}\label{KM}
\Gamma\,=\,-\frac{d\ln r}{dt}\,.
\end{equation}
Figure 2(b) shows the KM relaxation rate plotted as a function of time. At long times there is considerable noise, reflecting the fact that the relaxation rate is extremely sensitive to the $t\rightarrow\infty$ limit of the dielectric loss being determined accurately. Despite this noise it is clear that for the up and down
jumps ending at $174$ K the KM relaxation rates converge to the same number at long times. This shows that there is no so-called expansion gap, as Kovacs proposed in 1963 based on monitoring relaxation by measuring refractive index changes \cite{kov63}. The existence of an expansion gap was subsequently debated \cite{ren87,str97,mck99}, but Kolla and Simon recently concluded that there is no expansion gap \cite{kol05}. The present measurements confirm their simple picture that when equilibrium is approached, relaxation rates converge to a unique value. Note that the data of Fig. 2 contradict the widely used stretched-exponential relaxation function, $\exp[-(t/\tau)^\beta]$, because according to this function the relaxation rate approaches zero at long times, not a constant value.

\section{The ``inner'' clock}

Ageing experiments are usually interpreted utilizing the Tool-Narayanaswami (TN) formalism that models ageing in terms of an ``inner'' clock. This is based on the so-called material time, which is time measured on a clock with a clock rate that changes as sample properties evolve with time \cite{too46,hop58,nar71,sch86}. The material time may be thought of as analogous to the proper time of relativity theory, the reading on a clock following the observer's world line. The TN formalism is standard and widely used in industry for predicting ageing effects. But it is not known whether the material clock has real physical existence or is only a convenient mathematical construct. This question cannot be answered definitively, but the assumption that an inner clock does exist implies a definite prediction. Thus if a material time exists controlling all relaxations, the relaxation of the clock rate activation energy must be determined by the same material time that controls the dielectric ageing process. We proceed to formulate a quantitative, model-free test of this.

The dielectric loss of DBP has a high-frequency power-law decay described by $\epsilon''(f)\propto f^{-1/2}$ \cite{ols01,nie08}. Suppose the clock rate $\gamma(t)$ controls both structural relaxation and dielectric relaxation. We normalize $\gamma(t)$ such that it equals the loss peak frequency when the liquid is in equilibrium. Details are given in the Supplementary Information, but the main idea is illustrated in Fig. 2(c): Even out of equilibrium the dielectric relaxation rate $\gamma(t)$ may be determined by measuring the high-frequency loss utilizing

\begin{equation}\label{clockrate}
\epsilon''(f,t)\,\propto\,\left({\frac{\gamma(t)}{f}}\right)^{1/2}\,.
\end{equation}
If structural relaxation is also controlled by $\gamma(t)$, this expression allows us to determine the structural relaxation clock rate via 

\begin{equation}\label{structural}
\ln\gamma(t)\,=\,
2\ln \epsilon''(f,t) +A\,.
\end{equation}
The constant $A$ is found by calibrating using equilibrium data from higher temperatures where the loss peak is within the observable frequency range (Fig. 2(c)). 

The material time $\til$ is the actual time measured in units of the (time-dependent) inverse clock rate \cite {too46,ren87,str97,mck99,kol05,hop58,nar71,sch86}, $d\til=dt/\gamma(t)^{-1}$, i.e.  

\begin{equation}\label{mat_time_def} 
d\til\,=\,\gamma(t) dt\,.
\end{equation}
According to the TN formalism, for all temperature jumps the normalized relaxation function $R(t)\equiv r(t)/r(0)$ is some function of the material time that has passed since the jump, a function that is independent of the sign and size of the temperature jump: $R=R(\til)$. Generally the function $R(\til)$ depends on the quantity that is monitored.

The existence of an inner clock, i.e., the assumption that the dielectric clock rate equals the structural relaxation clock rate, can be checked without evaluating $\til$ explicitly or fitting data with analytical functions. This is done by proceeding as follows. First, define the dimensionless KM relaxation rate by

\begin{equation}\label{gammadim_def}
\tilde\Gamma\equiv \frac{\Gamma}{\gamma(t)}.
\end{equation}
This quantity is given directly by experiment since both $\Gamma(t)$ and $\gamma(t)$ are determined from $\ln\epsilon''(f,t)$. Next, we note that $\tilde\Gamma=-d\ln r/d\til=-d\ln R/d\til$. This implies that if $R(\til)$ is a unique function of $\til$, i.e., independent of sign and size of the temperature jump, then $\tilde\Gamma=\tilde\Gamma(\til)$ is also unique. In turn this implies that $\tilde\Gamma$ is a unique function of $R$:  

\begin{equation}\label{prediction}
\tilde\Gamma\,=\,\Phi(R)\,.
\end{equation}

Figure 2(d) shows that this prediction is fulfilled within the experimental uncertainty: The upper part of that figure plots the KM relaxation rates $\Gamma$ as functions of $R$ for the four temperature jump experiments of Fig. 2(a), the lower part plots $\tilde\Gamma$ as functions of $R$. The long time limit corresponds to $R\rightarrow 0$. The fact there is data collapse confirms the existence of a material time that controls both dielectric ageing and the ageing of the relaxation rate itself (the Supplementary Information provides mathematical details).

\begin{figure}
  \includegraphics[width=12cm]{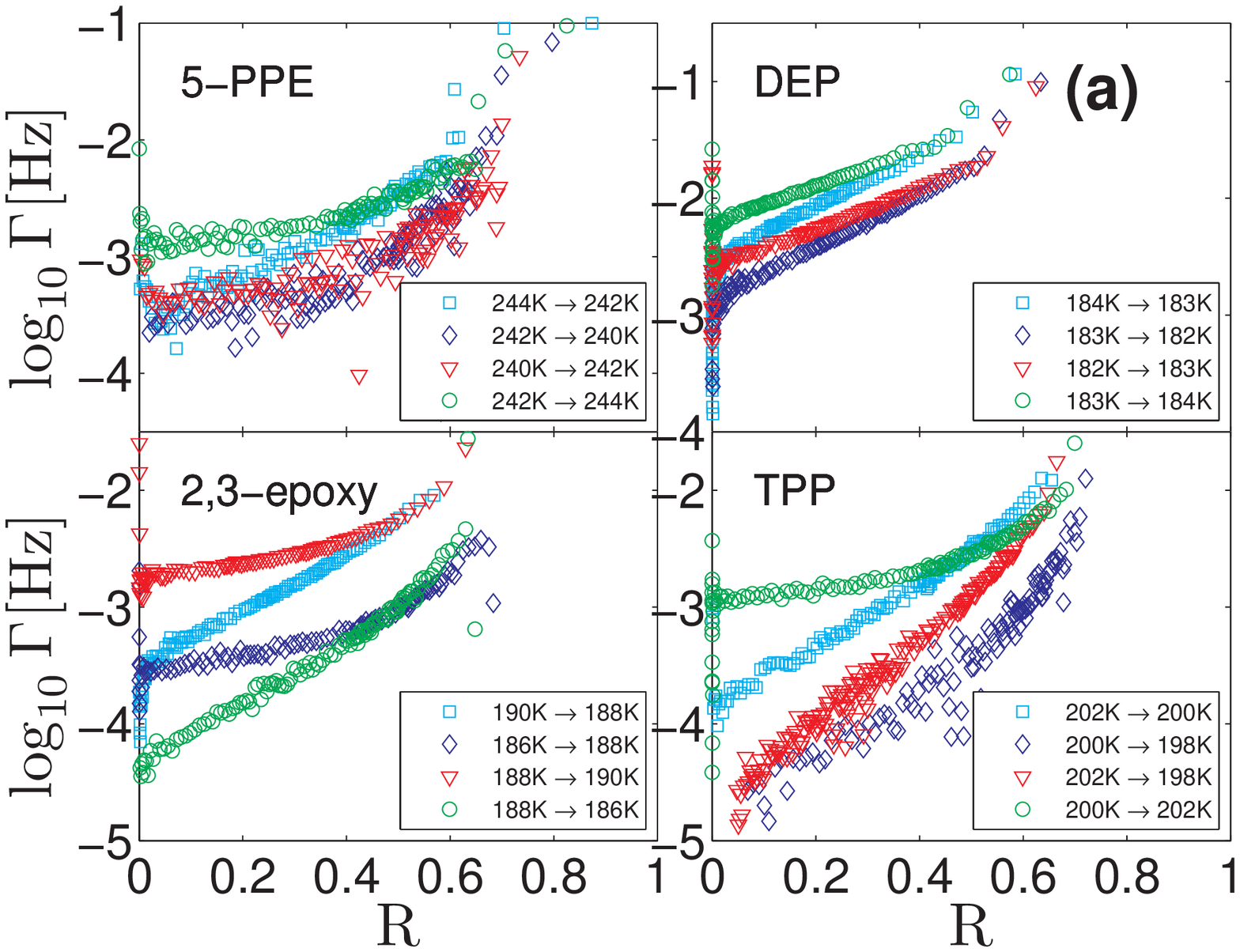}
  \includegraphics[width=12cm]{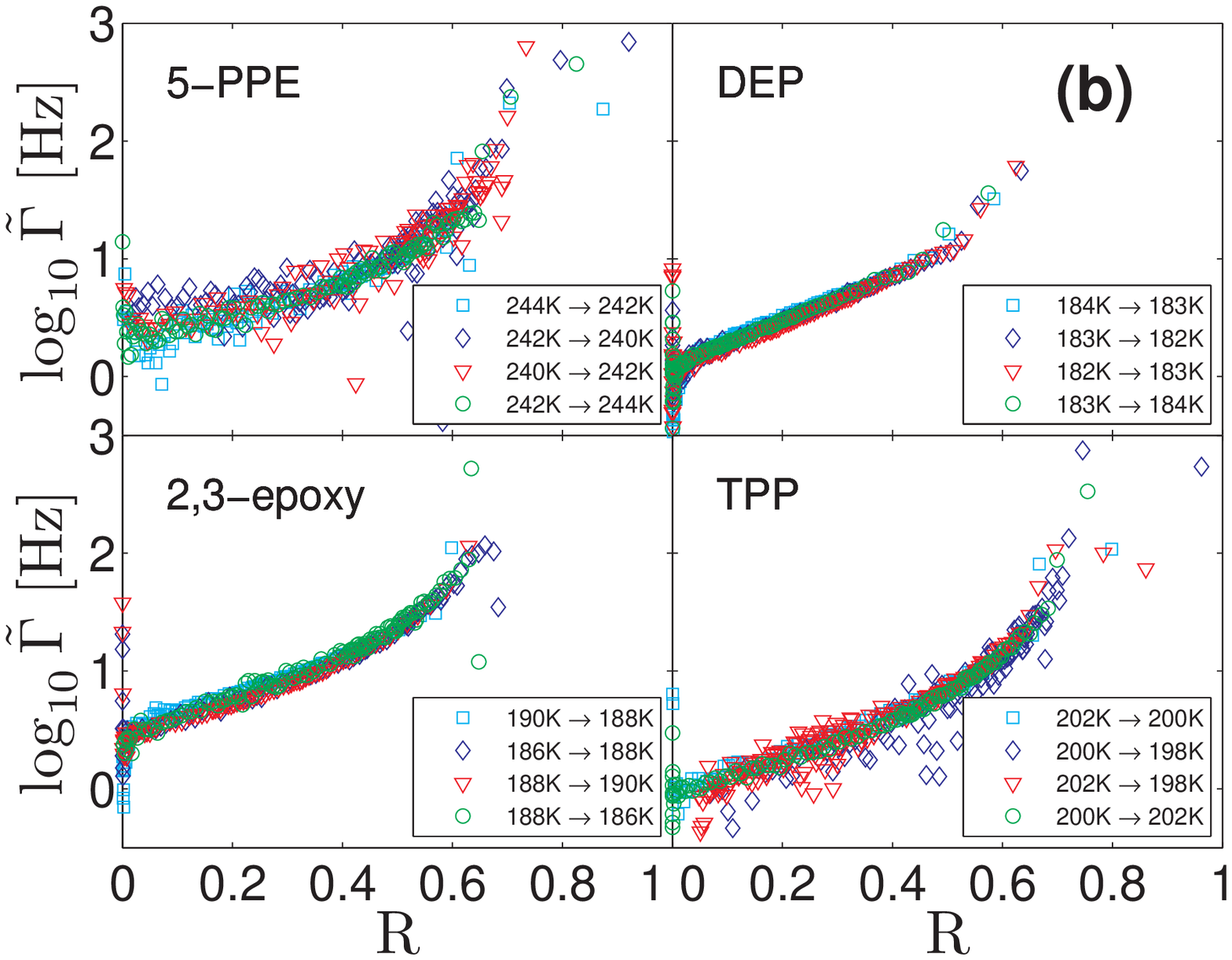} 
 \caption{Results for four other liquids. (a) Kovacs-McKenna relaxation rates as functions of the normalized relaxation function for the following four liquids: Di-ethyl phthalate (DEP); 5-poly phenyl ether (5PPE); 2,3 epoxyphenyl-propylether (2,3 epoxy); tri-phenyl-phosphite (TPP). (b) Dimensionless Kovacs-McKenna
   relaxation rates as functions of the normalized relaxation function for the same liquids. In all cases there is good data collapse within experimental errors, confirming the existence of an inner clock (a material time) for these liquids. Note that for TPP even the 4K down-jump follows the same master-curve as the 2K up- and down jumps. For all four liquids the dielectric loss was monitored at frequency $f=1$ Hz.}
\end{figure}

We repeated the experiment for four other liquids, diethyl phthalate, 5 poly-phenyl ether,  2,3 epoxy phenyl propylether and triphenyl phosphite. These liquids all have high-frequency dielectric losses proportional to $f^{-1/2}$ \cite{ols01,nie08}. The results for the KM rates $\Gamma(R)$ are plotted in Fig. 3(a) and the corresponding dimensionless KM rates in Fig. 3(b). The spread in KM rates at $R\cong 0$ reflects the already mentioned fact that relaxation rates are almost impossible to determine accurately at long times where the noise is of the same magnitude as the distance to equilibrium. In all cases we find good data collapse, confirming the existence of a material time for these liquids. 

\section {Details of the test}

We here detail the assumptions that must be made in order to arrive at the proposed model-free test of the material clock hypothesis. We first define a generally time-dependent dielectric relaxation rate $\gamma_d(t)$ by generalizing time-temperature superposition, then define the structural relaxation rate $\gamma_s(t)$, and finally arrive at the test.

\subsection {The dielectric relaxation rate defined for out-of-equilibrium situations}
The simple power-law high-frequency dielectric loss of DBP and the other liquids studied here, $\epsilon''(f)\propto f^{-1/2}$ \cite{ols01,nie08}, is utilized to monitor the dielectric relaxation rate $\gamma_d(t)$ as the structure ages following a temperature change by proceeding as follows. By definition $\gamma_d=f_p$ in the equilibrium liquid phase where $f_p$ is the dielectric loss-peak frequency. If temperature is lowered in a step experiment, the dielectric loss curve continuously moves to lower frequencies as the system relaxes to equilibrium. The idea is now (compare Fig. 2(c)) that how much the dielectric relaxation rate $\gamma_d$ has changed, i.e., how much the loss peak is displaced, may be determined from how much the loss at some fixed frequency in the high-frequency power-law region changes. Thus if the loss is continuously probed at a fixed frequency, the dielectric relaxation rate is determined from

\begin{equation}\label{1}
\epsilon''(f,t)\,=\, a (f/\gamma_d(t))^{-1/2}\,.
\end{equation}
Although this idea seems fairly straightforward, the concept of a dielectric relaxation rate has no obvious definition in a non-stationary situation as that of a relaxing structure. In order to specify the precise assumption needed to define $\gamma_d(t)$ and justify eq. (\ref{1}), we reason as follows: According to standard linear-response theory, for a system in thermal equilibrium the measured output is calculated from a convolution integral involving the history of the input previous to the measuring time. A convenient way to summarize time-temperature superposition for the equilibrium liquid is to formulate the convolution integral in terms of a dielectric ``material'' time $\til$: If $\gamma_d$ is the equilibrium liquid's dielectric relaxation rate ($\gamma_d=f_p$), the dielectric material time is defined from the actual time $t$ by $\til=\gamma_d t$. In terms of $\til$, if the input variable is the electric field $\bf E$ and the output is the displacement vector $\bf D$, the convolution integral is of the form

\begin{equation}\label{2}
{\bf D}(\til)\,=\,\int_0^\infty {\bf E}(\til-\til') \epsilon(\til')d\til'\,.
\end{equation}
Equation (\ref{2}) describes time-temperature superposition because it implies that, except for an overall time/frequency scaling, the same frequency-dependent dielectric constant is observed at different temperatures (we ignore the temperature dependence of the overall loss, an approximation which introduces an error of just 1\% over the range of temperatures studied).

In eq. (\ref{2}) the dielectric material time is defined from the actual time by scaling with $\gamma_d$, a quantity that is of course strongly temperature dependent. In the out-of-equilibrium situations following a temperature jump, the simplest assumption is that eq. (\ref{2}) also applies during the ageing process, but with the dielectric material time generalized by now assuming a time-dependent dielectric relaxation clock rate $\gamma_d(t)$:

\begin{equation}\label{3}
d\til\,=\,\gamma_d(t) dt\,.
\end{equation}
As $t\rightarrow\infty$ $\gamma_d(t)$ approaches the equilibrium liquid's loss peak frequency $f_p$ at the new temperature, of course. 

The equilibrium liquid's power-law dielectric loss $\epsilon''\propto f^{-1/2}$ applies whenever $f\gg f_p$. Equation (\ref{2}) thus implies that $\epsilon(\til')\propto\til'^{-1/2}$ whenever $\til'\ll 1$. It now follows that the proposed generalization of eq. (\ref{2}) now implies that the dielectric relaxation rate $\gamma_d(t)$ may be determined from of eq. (\ref{1}). 

\subsection{The structural relaxation rate}
We define the dielectric relaxation rate activation (free) energy $E$, which generally depends on time, by writing (where $\beta=1/k_BT$ and the prefactor $\gamma_0$ is taken to be $10^{-14}$ s)

\begin{equation}\label{4}
\gamma_d(t)\,=\,\gamma_0 \exp(-\beta E(t))\,.
\end{equation}
The activation energy depends on structure and it relaxes during strucural relaxation. From the Tool-Narayanaswami formalism one expects that after temperature changes the activation energy relaxes following a convolution integral over the temperature history, with a properly defined material time. For convenience we include the inverse temperature in the defining equation by writing  for small temperature variations $\Delta T(t)$

\begin{equation}\label{5}
\Delta (\beta E)(\til)\,=\,\int_0^\infty {\Delta T}(\til-\til') \phi(\til')d\til'\,.
\end{equation}
Here the structural relaxation material time is defined by 

\begin{equation}\label{6}
d\til\,=\,\gamma_s(t) dt\,,
\end{equation}
where $\gamma_s(t)$ is the rate of structural relaxation (one may have different structural relaxation rates and thus different material times for differing physical quantities relaxing (volume, enthalpy, ...)).

\subsection{Assuming the existence of a material clock}
The assumption that the material clock has real physical existence implies that the dielectric relaxation rate is identical to the structural relaxation rate:

\begin{equation}\label{7}
\gamma_s(t)\,=\,
\gamma_d(t) \,.
\end{equation}
This is not a trivial assumption. Thus Eqs. (\ref{2}) and (\ref{5}) may both well apply, but with different definitions of the reduced time. If Eq. (\ref{7}) is obeyed, however, Eq. (\ref{1}) implies that after a temperature jump the logarithm of the measured loss,

\begin{equation}\label{8}
\ln\epsilon''(f,t)\,=\,
\frac{\beta E(t)}{2}+C\,,
\end{equation}
relaxes with the rate that is determined from itself via Eq. (\ref{1}). It is this prediction that is tested in the main paper's Eq. (\ref{prediction}): The dimensionless KM relaxation rate $\tilde\Gamma$ of Eq. (\ref{gammadim_def}) is defined by dividing the KM relaxation rate $\Gamma$ by the structural relaxation rate $\gamma_s$; this quantity however is determined by means of Eq. (\ref{clockrate}) that actually gives the dielectric relaxation rate $\gamma_d$.

\section{Summary}

By measuring several relaxing quantities the existence of a material clock can easily be checked by simply investigating whether or not the quantities relax following the same function of time for various temperature jumps. This is a direct, model-free test, i.e., one that neither involves free parameters nor the fitting of data to some mathematical expression. For measurements of just one relaxing quantity as in this paper we believe that the above is the first model-free demonstration of the existence of an inner clock.

By modern micro engineering it should be possible to extend ageing experiment to even shorter times, thus making it realistic to perform a series of ideal temperature-jump experiments over just hours. Thus it is not unlikely that -- eventually -- ageing studies could become routine on par with, e.g., present-day dielectric
measurements.

\acknowledgments 
The centre for viscous liquid dynamics ``Glass and Time'' is sponsored by the Danish National Research Foundation (DNRF).

\end{document}